# The Intrinsic Properties of Brain Based on the Network Structure

Zou Xiang1, Yao Lie3, Zhao Donghua4, Chen Liang1, Mao Ying1,2,*

*1 Department of Neurosurgery, Huashan Hospital, Fudan Unuversity, Shanghai 200040, China*

*2 Tianqiao and Chrissy Chen International Institute for Brain Diseases, Shanghai 200040, China*

*3 Department of Pancreatic Surgery, Huashan Hospital, Fudan Unuversity , Shanghai 200040, China*

*4 Department of Mathematics, Fudan Unuversity, Shanghai 200433, China*

*Corresponding Author

E-mail: maoying@fudan.edu.cn

## INTRODUCTION

Brain is a fantastic organ that helps creature adapting to the environment. If we dissect functions from brain, there are at least three kinds of modules: receptor, effector and median integrated network. Receptor and effector is the most fundamental functions for a brain, even for a single neuron. In the advanced creatures, receptors can include some pretreatment process to integrate or distinguish information, and may affect our cognition to the world. **Tsien** et al.[1] has raised a postulate on the brain's basic wiring logic, indicating brain may organize the microarchitecture of cell assemblies as $2^n$-1 combinations that would enable knowledge and adaptive behaviors to emerge upon learning. Actually this hypothesis suggested the advanced receptor as a classifier. Effector in mammals is gifted more complex meanings such as emotion. All those responses from creatures are personal strategies for living.

The complexity of median integrated network varies among creatures. Nerve net is a very simple network in some coelenterate like polypus. Although is not considered as a brain, it can still perform between receptor and effector to make polypus curl up and adapt to live. For our human beings, many distinctive functions can be drawn like decision-making, memory, and more elusive appearance like consciousness. Decision-making process is actually one of the most pivotal functions. Intuitively, this process seems to be controlled by consciousness. But as we know, more decisions are made automatically in very short time subconsciously. Moreover, this process is not completely equal to conditioned response. Because on the one hand, making a quick decision does not rely on repeated experience. On the other hand, similar situation not always results in the same decision and vice versa. Learning is a modulation process for decision-making. By learning, we can gradually make better choices in certain situations. In order to describe this process, **Hebb** proposed that an increase in synaptic efficacy arises from a presynaptic cell's repeated and persistent stimulation of a postsynaptic cell. This theory does make sense, and is also proved in gastropod experience[2]. However, when take the network property into consideration, this theory may have some deficiency, which will be discussed in the following text.

Network is the most distinctive mesoscopic structure of brain. More details about this structure have been revealed over century after the findings by **Camillo Golgi**, but less understanding in-depth was achieved. Although we can rebuild a small brain by advanced imaging technique now days, functions can hardly be described. Or, the more microscopic structure we know, the more we will be confused. In spite of all those particulars, we want to verify capability of the simple network property. In this study, we try to expound some brain functions in a very basic view – the mathematical view, only with the structure of network, aiming to understand the intrinsic properties of brain.

## METHODS

### Mathematical model transformation

The simplest model of a neuron is a summator, which can sum up the input signal, judge and output a signal known as a spike. Even in this momentary process, thousands of reactions have happened, but all those details are ignored in this study except the most indispensable property. This kind of network model was firstly described by **McCulloch and W. Pitts**[3]. It is noteworthy that the neuro-behavior is discontinuous; neuron's output is represented by the density or probability of spikes, not the amplitude. Actually, the behavior of the neuro-network can be divided by a very short time interval, proposed as a spike unit. This is a sound approximate assumption, which will make the following study possible. Just as we consider, a real neuro-network is extremely complicated, such as different connections, synapse weights and thresholds are all unknowable. However, all those factors can be discretized and equivalent to other forms. Now we raise two equivalent and two discretized assumptions, aiming to transform the extremely complicated neuro-network behavior to a describable mathematical model (**Supplementary file**). Firstly, a real neuro-network can be imagined as a directed matrix (Figure 1A). The weight of each synapse $Wij$ means the $i$ to $j$ connection. As we know, pyramidal neurons are a kind of excitatory neurons, while interneurons are a kind of inhibitory neurons. So in Figure 1A, each row has the consistent sign of weight. In our mathematical model, we plan to break this limitation by type equivalence, which means an inhibitory neuron will be transformed to direct inhibitory synapses to the next neurons (Figure 1B). By this equivalence, every neuron in the matrix is similar, which has both excitatory and inhibitory synapses. The next is the threshold equivalence, in order to unitize each neuron's threshold of activation to zero (Figure 1B). Then synapse weigh discretization will unitize *Wij to* -1, 0 or 1, which will adjust to any weigh situation. Finally, the interval from signal transmission between neurons is also unitized, using a medium neuron to make delay. Figure 1C is a given network that expressed by a directed matrix, then transformed to an equivalent simplified matrix. In theory, this equivalent simplified matrix can describe any kinds of complex network, as long as there are enough elements. This mathematical model seems to make a natural network become more complicated. Actually, this model has the maximal freedom, which will reveal any possible behavior of a natural network. Thus, we call this mathematical model a differential of neuro-network.

**Operation rules of the mathematical model**

The dynamics of the neural network can be expressed by the following recursive formulas:

$$M = \begin{pmatrix} m_{11} & \cdots & m_{1l} \\ \vdots & \ddots & \vdots \\ m_{k1} & \cdots & m_{kl} \end{pmatrix}, k = l\,,\ m_{kl} \in \{-1,1\} \quad (1)$$

$$S(t) = (Y_1^t\ Y_2^t\ \dots Y_i^t\ )\ ,\quad Y_i^t \in \{0,1\} \quad (2)$$

$$X\ = S(t-1) \times M \quad (3)$$

$$Y_i^t\ = F[X_i^t] = \begin{cases} 0, & when\ X_i^t \le 0 \\ 1, & when\ X_i^t > 0 \end{cases} \quad (4)$$

$$S(\mathrm{t}+1) = F[S(t) \times M] = (Y_1^{t+1}\ Y_2^{t+1}\ \dots Y_i^{t+1}\ )\ , \quad (5)$$

*M* is the equivalent simplified matrix mentioned above. *S(t)* is the state of the present network, assembled by $Y_i{}^t$, means the activation (1) or silence(0) of one neuron. *X* is the medium process before judgment by function *F*, deciding its activation or silence at the next time point. Although we now have little prediction of the behavior or dynamics about this equations set, it is worth noting that, the issue we concerned about is not one state of $Y_i{}^t$ set, but the distribution of $Y_i{}^t$ over time, which is known as spikes.

## RESULT

### General features of the simplified neuro-network

Firstly, we assume that the basic function unit in a brain is neural assembly. A certain assembly is relatively closed, with an appearance of random inner connections, which is our study focus. In contrast, connections among assemblies are more likely to be nonrandom wirings, organizing function units in a logical construction, which will be studied further. Although not be proved, we are more likely to believe that the meaning of spikes lies on its probability or frequency. This is the premise to understand the language of neurons.

If there is no interference to an assembly, it is not difficult to understand that the amount of $Y_i{}^t$ set is finite, relying on the previous state of the network known as $S(t-1)$, meaning the state of the network is fixed point or a loop. This process is very similar to the **Markov process**, but also with fundamental differences. Actually, as the internal/external environment will affect the state, a closed neuro-network is difficult to maintain in creatures, leading to a non-repeating dynamic behavior, even if there is no synapse remodeling.

In a closed network constructed by random synapse, we have already proved that the spike probability distribution of each neuron is in accord with a sigmoid curve in the assembly (**Supplementary file**). In the rest stable state, the total assembly activity is related to the proportion of excitatory synapse. The spike probability ($P_i$ ) of a certain neuron has weak correlation with the expectative accumulation calculated by connection matrix, but has close relation to the posterior average input accumulation $\overline{X_i}$. The relationship can be expressed as the following equation:

$$P_i = \frac{1}{e^{A\overline{X_i}+C}}$$

$P_i$ is the spike probability, $X_i$ is the expected accumulated input, A and C are

network related parameters. This is the basic equations describing the relationship between spike probability and average input accumulation for a neuron. In Figure 2A, we demonstrate a representative example for a random connective 1000 size neural assembly, with the similar amount of excitatory and inhibitory synapse. Spikes in time series for each neuron are shown in Figure 2B, which are all with a fixed activation probability but without any regularity of distribution. When taking average accumulation (average input signals on each time point) into consideration, the relationship presents as a sigmoid curve (logistic function, Figure 2C). It is worth noting that this is a closed network without any affection, the sigmoid curve is automatically formed by a proper excitatory/inhibitory proportion.

**Spike probabilities in the assembly can meet the solution of a nonlinear equation set**

It is for certain that the activation pattern of each neuron is decided by the connection pattern of the network, which is defined as directed connection matrix *M* as mentioned above. However, it is impossible to predict the spike probability of each neuron directly by *M*. The reason is that the spike probability is the solution of a non-liner equations set, and has been proved (**Supplementary file**). In particular, we cannot calculate the spike probability by sum the synapses towards one neuron in *M*, even for the expectation. In Figure 3A, we give a 500 closed neuron cluster, the left panel showed the origin activation pattern of each neuron. However, when we adjust matrix *M* by summation of total synapse to each neuron then array in ascending or descending order, the spike probability is not strictly arranged by the same order, despite of the general trend. In a 1000 closed neuron cluster, spike probability of each neuron has weak correlation with the expectative accumulation from the network (Figure 3B). In addition, when stimulants exist, the strength of continuous stimulation to one neuron has weak correlation with its spike probability shift (Figure 3C). In most situations, the activation pattern will not change synchronously only by swapping the row in *M* (Figure 3D). The entire phenomenon mentioned above indicated that the spike probability of one neuron in the assembly is not a solution of a simple or linear equation. Here is an equation set to describe their relationship:

$$S_i + (P_1 \cdots P_i) \times M = \frac{1}{A} \cdot (\log\frac{1-P_1}{P_1} - C \cdots \log\frac{1-P_i}{P_i} - C)$$

In this equation set, $S_i$ is the summation of external stimulation to one neuron in the neural assembly. If there is no stimulant, $S_i$ equals to zero. Generally, to solve the equation, the external stimulation is relatively unchanging in a short time interval, till the spike probability shift becomes stable.

**Distribution shift of the spike probability under stimulations**

Till now, we only describe the dynamic patterns in a closed network. In the changing environment, the network will be affected by inputs every moment, leading to a more certain state in usual. We have already proved that, in a neural assembly, the increasing stimulation strength will elevate the slope of the original sigmoid curve, as well as polarize the spike probabilities of each neuron, leading to a lower Boltzmann entropy (or Shannon entropy, **Supplementary file**). Figure 4A showed a closed neuro-network that received inputs by step. We can see the spike probabilities of each neuron shifted by the increasing input. Entropy as well as the amount of active neuron of the whole assembly was decreased by steps. Some spike probabilities were raised up while some were suppressed, which were polarized (Figure 4C). As mentioned above, the changing rules were in accord with the solution of equation. This phenomenon has an obvious implication, meaning the more information we get, the more convinced we make a choice.

**The stability of the simplified neuro-network**

The stability of a neuro-network is of great of importance for its functions. To be adaptive to the changing environments, the network transforms to a certain stable state, meaning the activation of each neuron switches to a certain mode, being excited or inhibited. In this simplified assembly network, the stability is closely related to the distribution of -1 and 1 in matrix *M*. The simulation test also demonstrated that the amount of neuron is another factor affects the network's stability. Even the spatial distribution of -1 and 1 is set by random; stability seems to be an intrinsic property under proper parameters. Figure 5A shows the network's stability under different parameters. In order to keep the activation of each neuron stable in a closed network, the distribution of -1 and 1 in the $M$ should be near half and a half to avoid over excited or inhibited (Figure 5B). As the neuron number increasing, the proportion of 1 has to grow in step to keep stable. In the real brain, we have already known that the importance of excitatory and inhibitory distribution. Disordered distribution will lead to various neurological diseases. It is notable that the proportion of 0 in matrix M will not influence the stability in a closed network, which is related to synapse density (Figure 5C). However, when the network is not closed, means the existing extra-input, the network behavior will differ among various synapse densities.

In a stable closed neuro-network, we have already proved the total spike rate is related to the amount of neuron as well as the proportion of excitatory and inhibitory neurons (**Supplementary file**). The details can be described by the function:

$$N(t) = N \cdot \left[ \Phi\left( \frac{N(t) - N(t) \cdot E_1}{\sqrt{N(t) \cdot E_1 \cdot E_{-1}}} \right) - \Phi\left( \frac{\frac{N(t)}{2} - N(t) \cdot E_1}{\sqrt{N(t) \cdot E_1 \cdot E_{-1}}} \right) \right]$$

In this transcendental equation, $N(t)$ means the expected amount of active neurons in one time point. $N$ is the total neuron amount, while $E_1$ and $E_{-1}$ are the excitatory and inhibitory neuron proportions. $\Phi$ is the accumulation density function for normal distribution. Figure 5D demonstrated some randomly connected neuron assemblies (500 neurons) with various excitatory neuron proportions, which were in accordance with our prediction.

**How can memory be formed without synapse plasticity: a hypothesis**

Since the spike probability of each neuron in the assembly only depends on the connection matrix and stimulant, it seems that memory will only be stored in synapses. However, the phase of firing will be labeled in the spike sequence after stimulation, making it possible for assemblies coupling. According to the equation set, a high strength stimulant will push one neural assembly to a deterministic state (polarized spike probability), thus holding the phase information. When the stimulant disappears (or switches to another stimulant), neurons in this assembly will fire in another sequence starting from a certain phase.

As in a single neural assembly, both spike probablties and relative phase among neurons are decided by the stimulation pattern, so neurons in one assembly will not be coupled freely. However, neurons from different assemblies can achieve free couping. For example, 2 neurons with 0.25 and 0.5 homogeneous spikes probability, can coulpe to a downstream neuron with two major patterns (synchronous or asynchronous). With different threholds, the downstream neuron will fire on vaiours probablity range from 0.0 to 0.9 (Figure 6A). Another desciption about neurons coupling is avaliale in the **Supplementary file**. Figure 6B showed a framework of a simple network that can achieve memorizing. The goal of the framework is to distinguish the past inputs by modulate the threshold, when the inputs is dismissed. Firstly, the two different inputs are connected to the coupling assemblies, which are independed to each other to make coupling possible. Then four coupling units are connectetd to the output unit that is also unidirectional. As mentioned above, high or low threshold will lead to different coupling results, so there is also a threshold modulator affects the output signal. Our results showed that for diiferent input modes, the output can demonstate obvious distinguishable fring patterns, with the influence of step-up threshold (Figure 6 C and D). Here we only foucus on the spike probability conbination as pattern, not the detailed firing sequence. Depends on the theory above, we build a hypothetical cognition-memory system, which can be an analogy of holography. Attention is a coden signal pattern acting on this system for memory store and recall (**Supplementary file**).

## DISCUSSION

In this study, we constructed a simplified neuro-network to simulate a neural assembly, and found the rules of spike probability distribution by random connection, finally inferred an equation set in which the solutions can meet the spike probabilities of each neuron.

The simplification of the real neural assembly in this study is based on an unattested but straightforward logic: a more complex network cannot be worse than a simplified network. Thus, the real neuro-network will contain all the functions of a simple one, enable our findings to be a small piece of the whole.

According to our findings, the general activity of a neural assembly is closely related to the excitatory/inhibitory proportion. Although the parameters in the equation are only the assembly size and excitatory/inhibitory proportion in the connection matrix, it can also be realized that other factors like threshold will contribute to the network activity. Those factors can be equivalent to the excitatory/inhibitory proportion as mentioned above. In creatures, brain can adjust the whole assembly activity by micro-environment such as blood supply, hormone and neurotransmitter, but those factors are not included in this study.

In creatures, it is not easy to represent a certain number in any shape or form, let alone perform operations by numbers. Thus, 'all or none' is the less-than-ideal alternative for brain to represent number. By frequency or probability of spikes, brain can even perform operations and give solutions. The equation set indicates brain can have the mathematics property, appears as receiving digitized signal and then transform the digital result to responses. This is the intrinsic function of the neural assembly. This function can be transformed to many other equivalent forms to solving problems. As we can see, synapse connection is the infrastructure of network dynamics. At the same time, stimulus is another important factor that contributes to the spike probability in a short duration. This relationship is just like the function and variables. In long term duration, network state will modulate the connections, thus change the solutions for a certain problem. We can foresee that this mechanism is far more that Hebb rule, on account of the nonlinear property of the equation set.

The theoretical foundations for modern research on decision-making were laid within the development of evidence accumulation models[4,5]. Experiments in vivo revealed the accumulator value is closed related to the decision probability or firing rate, presented as a sigmoid curve. According to our findings, firing rate distribution in a neural assembly can spontaneously form a sigmoid curve, with or without stimulation.

That is to say, decision-making is the natural ability for living. It has already been proved that logistic function is kind of maximum entropy distribution, conforming to the decision-making principle. Under the restriction of synapse, the maximum entropy distribution is the most proper distribution of spike probability that with minimum miss-selection risk. For instance, in the decision-making process, we will make a decision by random without any useful information unless there is sufficient reason to make a choice.

Although we proved that a random simplified neuro-network has the intrinsic spike probability distribution principle, most of our brain are not wiring by random, at least not in all scales. Wiring pattern is the most significant signature that controls the neuro-behavior, but the connective rules cannot be understood directly. The causality from wiring to spikes is not linear, which is indicated to be the solution of a linear function set. Although connections in a certain assembly seem to be random, connections among assemblies are more likely to be nonrandom wirings, organizing function units in a logical construction.

For a certain neural assembly, when there's no input, ground state is stable and determined by inner connections. Each specific input will motivate the assembly to another state rapidly, which is unique and meaningful, representing a solution of the equation mention above. The transition period between continual states is unstable and meaningless. Actually, there is no long lasting state due to the changing external stimulus. In a relatively stable network state, each neuron has a fixed frequency or spike probability. Spike frequency (probability) of one neuron is the definite and meaningful solution under a certain stimulant. Shifting frequency of one neuron constitutes to heterogeneous firing sequence, which is common in the live recording. We considered that a more stable spike frequency has the representational meaning, rather than a heterogeneous sequence. Spikes are recognized as frequency, and coupled by sequence.

To be precise, the entire hypothesis in this study can only describe what happened in a real brain by seconds time scale. Although the intrinsic dynamics of this simplified network have explained some neural functions, our brain can have far more dimensions interlace than network. For instance, synapse remodeling, hormonal regulation and neuron subtypes are all critical for complex functions in a broader spatial and temporal scale.

To summarize, the value of this study is as follows. Firstly, we mathematically prove the intrinsic character of a proper network, that is to say the spike probability distribution is a spontaneous phenomenon, indicating the decision making mechanism. Secondly, we also proved the spike probability distribution is a maximum entropy distribution under one certain stimulus, indicating the nature of intelligence. Thirdly, we propose the importance of phase under a certain spike probability distribution, and then hypothesize the 'dominant effect', indicating the mechanism of short-term memory.

**Figure Legends**

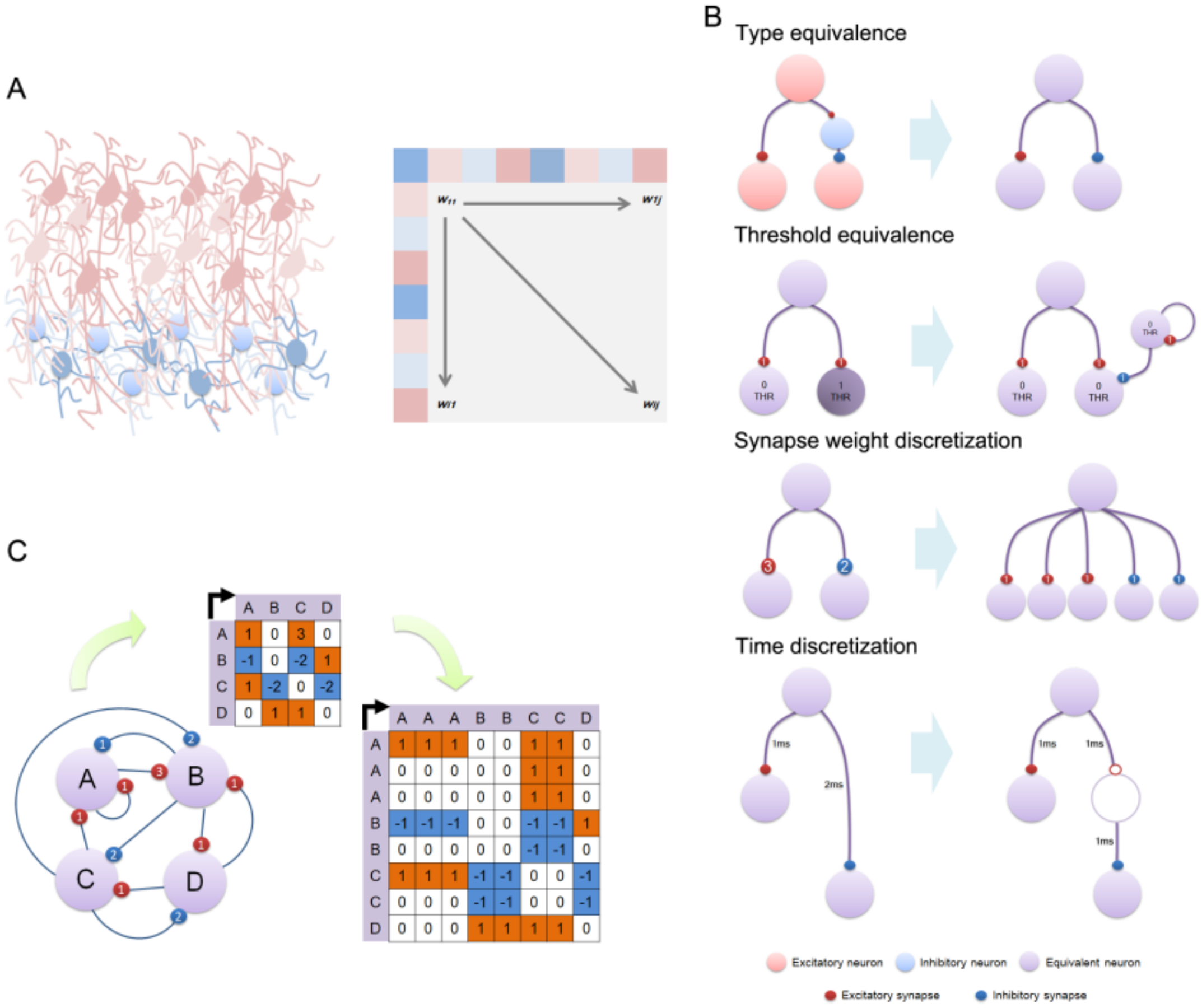

**Figure 1. Equivalence and discretization of the connection matrix.** (A) illustration of a connection matrix from a neural assembly. (B) two equivalence and discretization ways for the connection pattern. (C) illustration of the equivalence and discretization for a connection matrix.

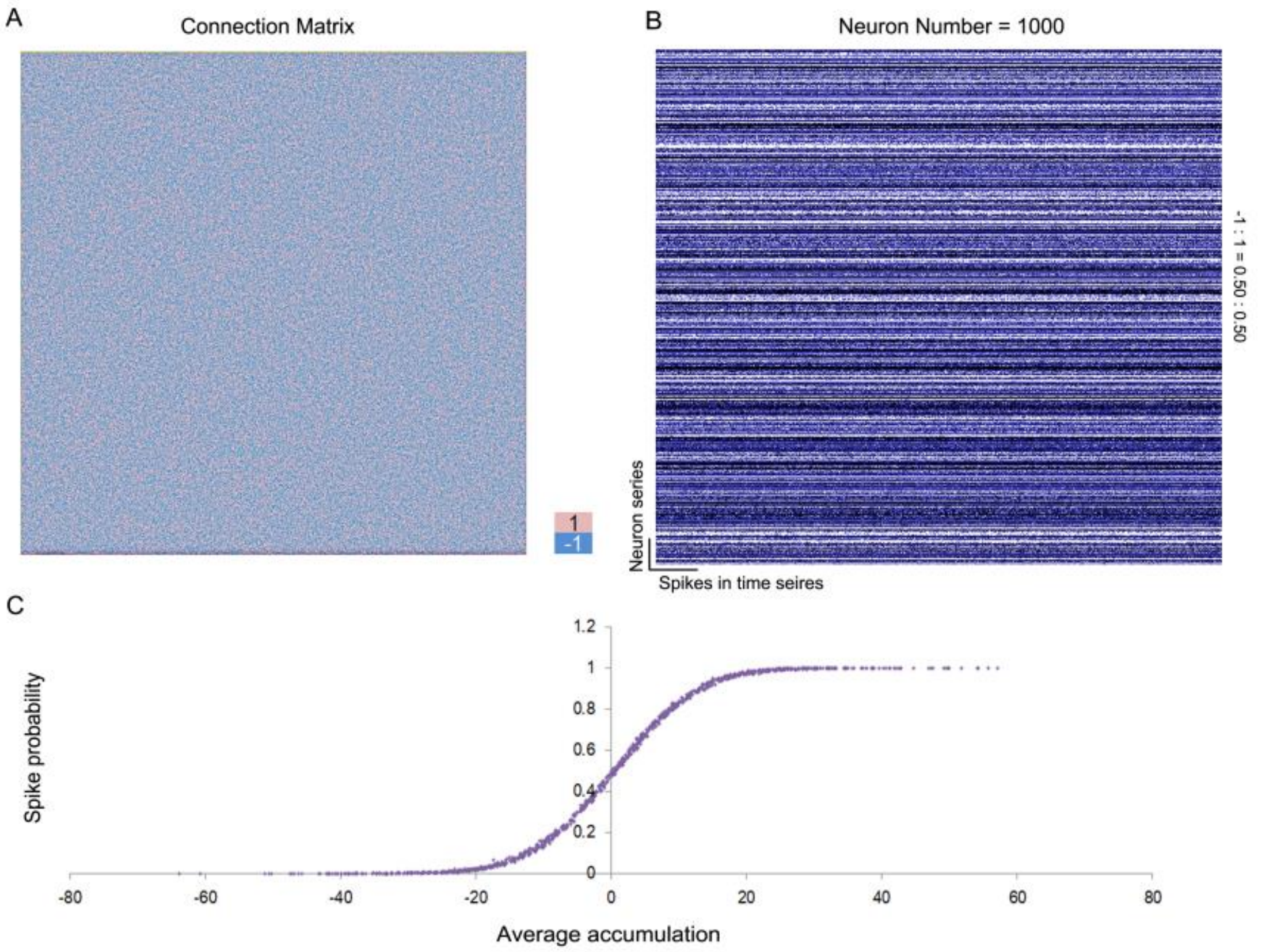

**Figure 2. Spontaneous spike probability distribution of a closed neural assembly.** (A) random connection matrix with a similar excitatory/inhibitory proportion. (B) spike-time series of each neuron in the assembly. (C) average accumulation-spike probability curve of the neural assembly.

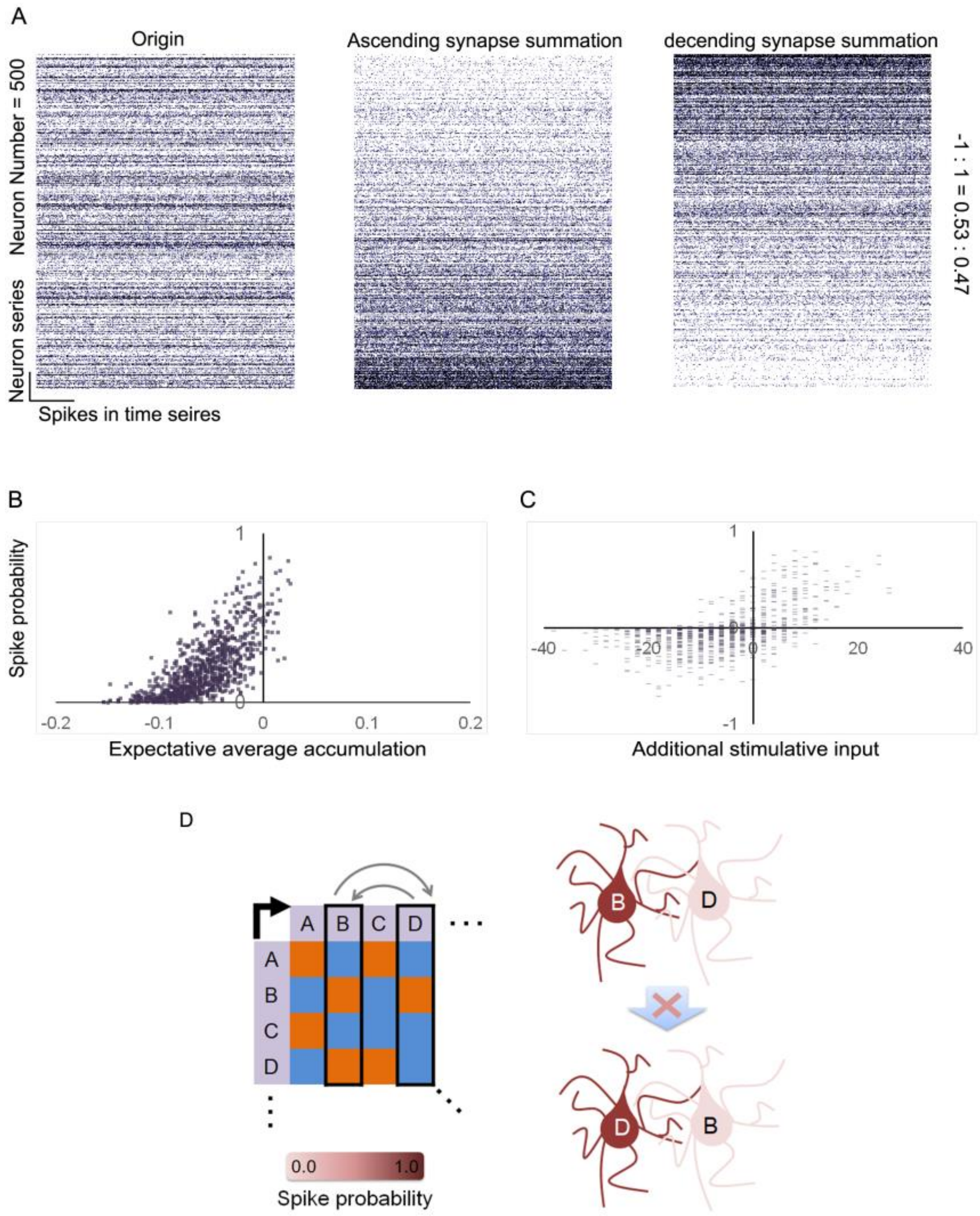

**Figure 3. Spike probabilities cannot be predicted by linear operations.** (B) weak correlation between the actual spike probability and the expected probability calculated by synapse weight summation. (C) weak correlation between the spike probability shift and the expected shift calculated by input summation. (D) swopping column of the connection matrix cannot maintain the previous spike probability.

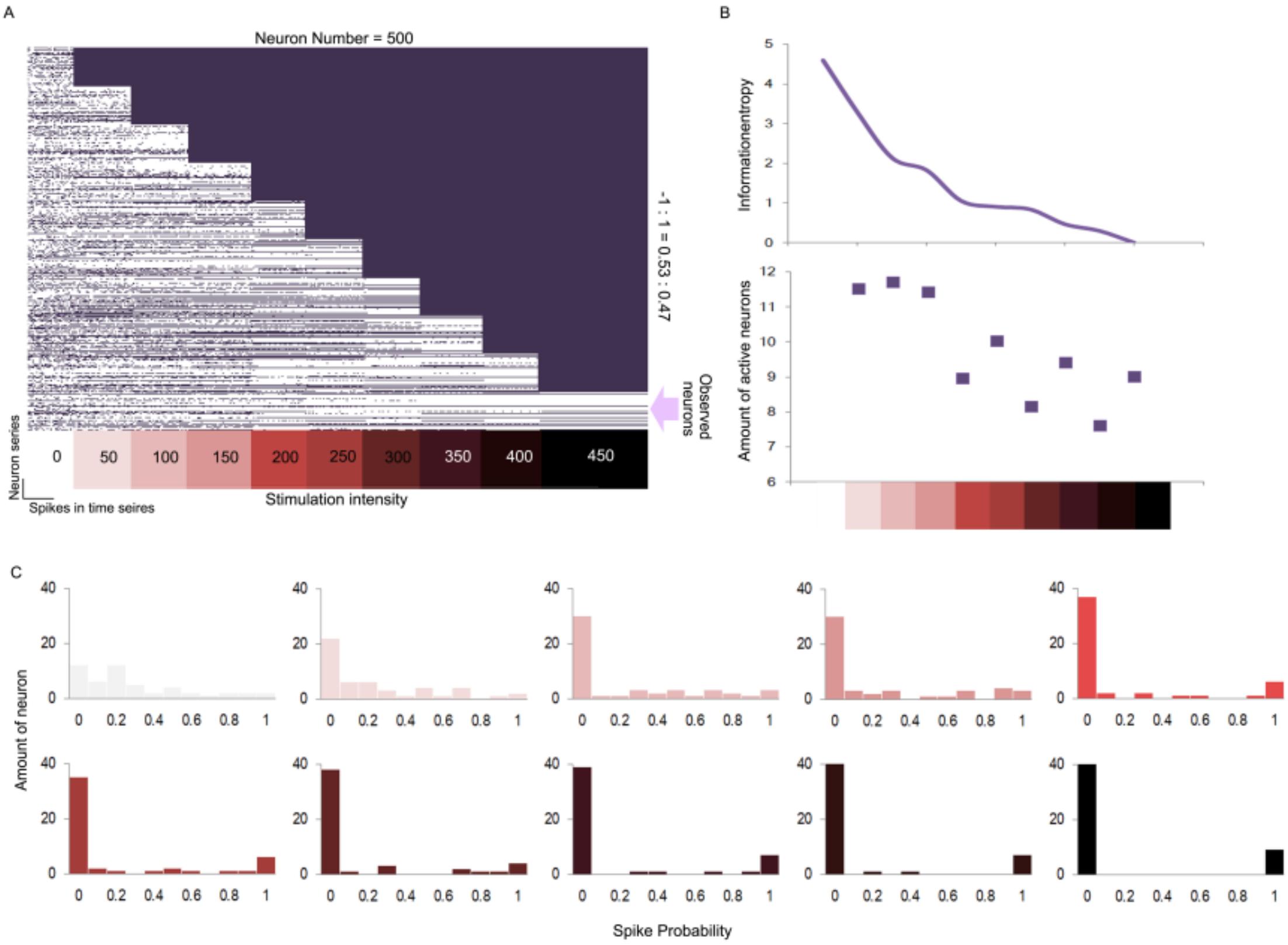

**Figure 4. Polarization of the spike probability by increasing stimulations.** (A) spike-time series of each neuron in the assembly, affected by increasing stimulations. (B) information entropy of the observed neuron spike probabilities by increasing stimulations. (C) polarization of the spike probability by increasing stimulations.

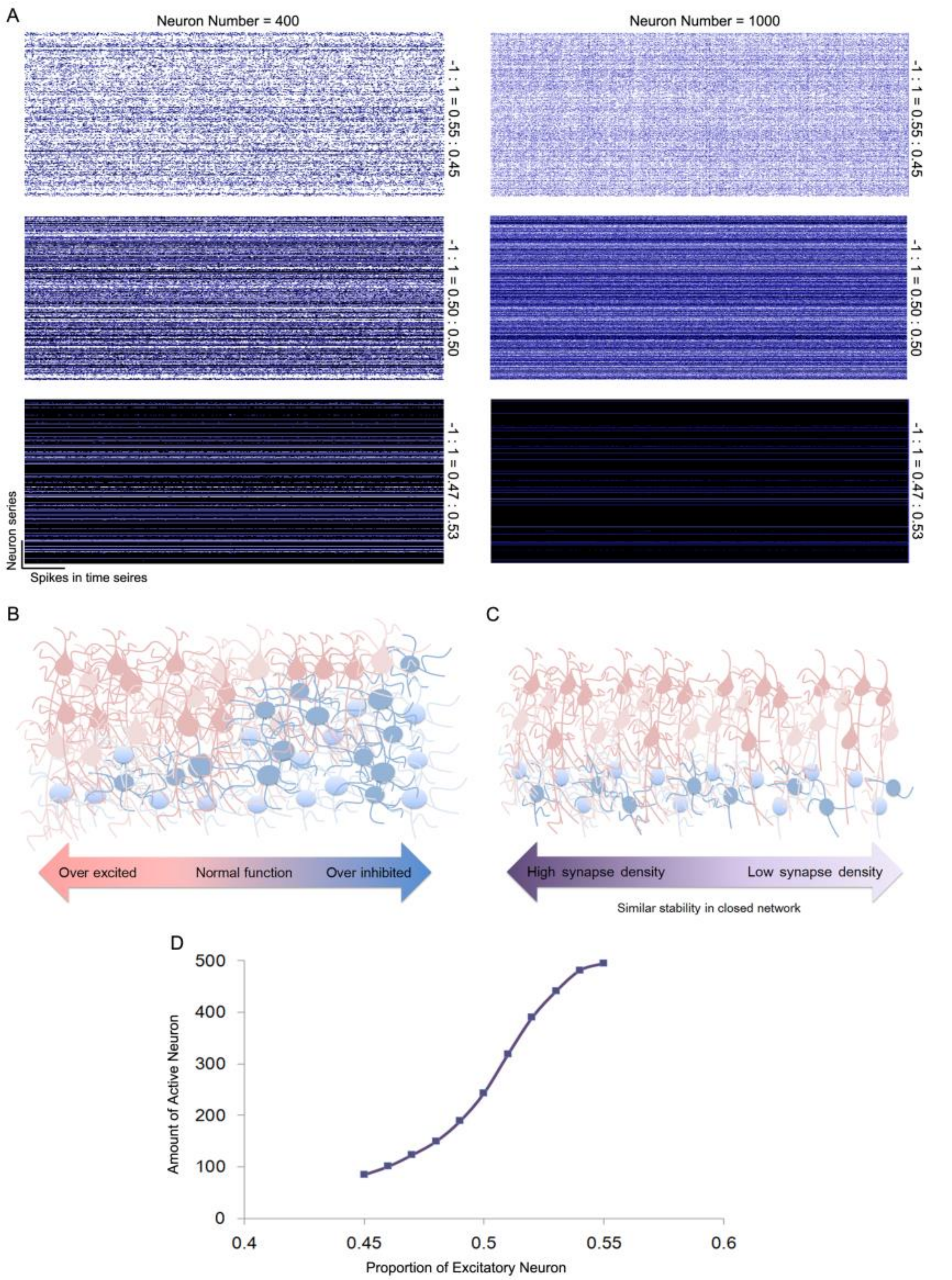

**Figure 5. Activity of a closed neural assembly depends on the neuron amount and excitatory/inhibitory proportion.** (A) spike-time series of each neuron in the assembly with different neuron amount and excitatory/inhibitory proportion. (B) and (C) illustration diagram of the activity regularity with different parameter. (D) Relationship between excitatory/inhibitory ratio and assembly activity.

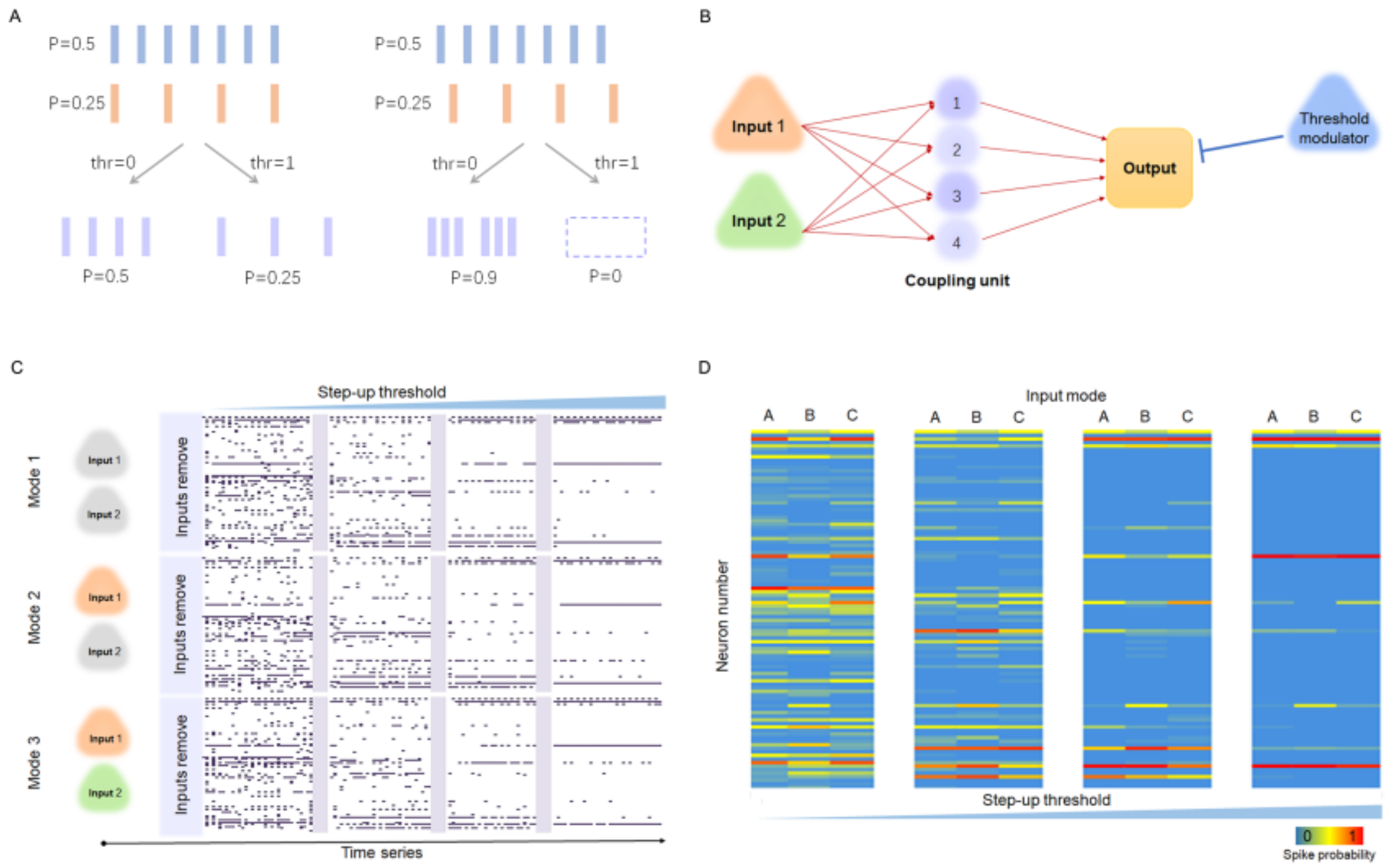

**Figure 6. Memory property of the neural assembly without synapse plasticity.** (A) different coupling phase and threshold leading to different output from two same neurons. (B) connection framework of neural assemblies embody memory. (C) spike-time series of the outcome by three phase coupling modes, under an increasing coupling threshold. (D) heat maps of outcome spike probabilities by three phase coupling modes, under an increasing coupling threshold.

**Supplementary file**

**Proof 1**

$$M = \begin{pmatrix} m_{11} & \cdots & m_{1l} \\ \vdots & \ddots & \vdots \\ m_{k1} & \cdots & m_{kl} \end{pmatrix}, k=l,\ m_{kl} \in \{-1,1\} \quad (1)$$

$$S(t) = (Y_1^t\ Y_2^t\ \dots Y_i^t\ ) \ ,\ \ Y_i^t \in \{0,1\} \quad (2)$$

$$X\ = S(t-1) \times M \quad (3)$$

$$Y_i^t = F[X_i^t] = \begin{cases} 0,\ when\ X_i^t \leq 0 \\ 1,\ when\ X_i^t > 0 \end{cases}, \quad (4)$$

$$S(\mathrm{t}+1) = F[S(t) \times M] = (Y_1^{t+1}\ Y_2^{t+1}\ \dots Y_i^{t+1}\ )\ , \quad (5)$$

As the ensemble $X_i^t$ and $Y_i^t$ is limited as a circle T, by a large number of dimension *i* as well as iterations, $X_i^t$ will form a distribution. For each $X_i^t$ , $\overline{X_i^t}$ can be calculated by $\sum_1^T \frac{X_i^t}{T}$. During iterations, the probability of $Y_i^t = 0\ or\ 1$ can also be calculated as $P_i^0\ or\ P_i^1$. The relationship between $\overline{X_i^t}$ and $P_i^1(or\ P_i^0)$ is expressed as follows.

By enough iterations, means $t \to \infty$, $\overline{X_i^t}$ will be fixed as $\overline{X_i}$, has less relationship with t. Define an ensemble $X = \{\overline{X_1}\,\overline{X_2} \cdots \overline{X_i}\}$. Regarding that $Y_i^t = 0\ or\ 1$ is like some kind of sampling process from ensemble $X$, each distribution can be considered as normal distribution by sufficient large of *i* and *t*, as:

$$P(\overline{X}|Y_i^t = 1) \sim N(u_1,\ \sigma^2)\ , \quad (6)$$

$$P(\overline{X}|Y_i^t = 0) \sim N(u_0,\ \sigma^2)\ , \quad (7)$$

According to Bayes theorem,

$$P(Y_i^t = 1|\overline{X}) = \frac{P(\overline{X}|Y_i^t = 1) \cdot P(Y_i^t=1)}{P(\overline{X})}\ , \quad (8)$$

$$P(Y_i^t = 0|\overline{X}) = \frac{P(\overline{X}|Y_i^t = 0) \cdot P(Y_i^t=0)}{P(\overline{X})}\ , \quad (9)$$

So,

$$\log\frac{P(Y_i^t=1|\overline{X})}{P(Y_i^t=0|\overline{X})}=\log\frac{P(\overline{X}|Y_i^t=1)\cdot P(Y_i^t=1)}{P(\overline{X}|Y_i^t=0)\cdot P(Y_i^t=0)}$$

$$=\log P(\overline{X}|Y_i^t=1)-\log P(\overline{X}|Y_i^t=0)+\log\frac{P(Y_i^t=1)}{P(Y_i^t=0)}$$

$$=-\frac{(\overline{X}-u_1)^2}{2\sigma^2}+\frac{(\overline{X}-u_0)^2}{2\sigma^2}+\log\frac{P(Y_i^t=1)}{P(Y_i^t=0)}$$

$$=\frac{2(u_1-u_0)\overline{X}+({u_0}^2-{u_1}^2)}{2\sigma^2}+\log\frac{P(Y_i^t=1)}{P(Y_i^t=0)}$$

So,

$$\frac{P(Y_i^t=1|\overline{X})}{P(Y_i^t=0|\overline{X})}=\frac{P(Y_i^t=1|\overline{X})}{1-P(Y_i^t=1|\overline{X})}=e^{\frac{2(u_1-u_0)\overline{X}+({u_0}^2-{u_1}^2)}{2\sigma^2}+\log\frac{P(Y_i^t=1)}{P(Y_i^t=0)}}$$

So,

$$P(Y_i^t=1|\overline{X})=\frac{1}{1+e^{\frac{(u_0-u_1)\overline{X}}{\sigma^2}+\frac{({u_1}^2-{u_0}^2)}{2\sigma^2}+\log\frac{P(Y_i^t=0)}{P(Y_i^t=1)}}}\sim\frac{1}{1+e^{A\overline{X}+C}} \tag{10}$$

Easily know that A<0, C is constant. $P(Y_i^t=1|\overline{X})$ conforms to a logistic curve.

So the relationship between $\overline{X_i^t}$ and $P_i^1$ is $P_i^1=\frac{1}{e^{A\overline{X}+C}}$.

Besides,

$$\overline{X_j^t}=\frac{\sum_1^T X_j^t}{T}=\frac{\sum_1^T\sum_1^i(Y_i^t\times m_{ij})}{T}=\frac{\sum_1^T\sum_1^i Y_i^t}{T}\times m_{ij}=\frac{\sum_1^i\sum_1^T Y_i^t}{T}\times m_{ij}$$

$$=\sum_1^i P_i^1\times m_{ij}=(P_1^1\ P_2^1\cdots P_i^1)\times M$$

Bring this function to (10),

$$(P_1^1\ \cdots P_i^1)\times M=\frac{1}{A}\cdot(\log\frac{1-P_1^1}{P_1^1}-C\cdots\log\frac{1-P_i^1}{P_i^1}-C)$$

**Proof 2**

Assuming that there is an iteration system described by function (1)-(5). For matrix (1), we can set the parameters $E_{-1}$ and $E_1$, as well as $N^2$, representing the expected proportion of -1 or 1, and the amount of the elements ($E_{-1} + E_1 = 1$).

As mentioned above, when $t \to \infty$, The state of vector $S(t) = (Y_1^t \ Y_2^t \ ... Y_i^t\ )$ will be trapped in a circle T ($Y_i^t \in \{0,1\}$). The probability of $Y_i^t = 1$ ($P_i^1$) is the solution of nonlinear system of equations (11):

$$(P_1^1 \cdots P_i^1) \times M = \frac{1}{A} \cdot (\log \frac{1-P_1^1}{P_1^1} - C \cdots \log \frac{1-P_i^1}{P_i^1} - C) \quad (11)$$

Let us define a value $\widetilde{N}$, means the expected amount of $Y_i^t = 1$ by each t. Here we doubt that if $\widetilde{N}$ is stable and how to calculate $\widetilde{N}$? The proof is as follows.

Supposing that at one time t, the amount of $Y_i^t = 1$ is $N(t)$, and $N(t) < N$. Bring (3) to (4), the value of $Y_i^{t+1}$ is determined by:

$$Y_i^{t+1} = F[X_i^{t+1}] = \begin{cases} 0,\ when\ (Y_1^t \ Y_2^t \ ... Y_i^t\ ) \times \begin{pmatrix} M_{1i} \\ \vdots \\ M_{ki} \end{pmatrix} \leq 0 \\ 1,\ when\ (Y_1^t \ Y_2^t \ ... Y_i^t\ ) \times \begin{pmatrix} M_{1i} \\ \vdots \\ M_{ki} \end{pmatrix} > 0 \end{cases}, \ i = k$$

Taking the parameters $E_{-1}$ and $E_1$ into consideration, we can easily see that when $E_{-1} < E_1$, the probability of $Y_i^{t+1} = 1$ is higher than the situation $E_{-1} > E_1$. That is because $(Y_1^t \ Y_2^t \ ... Y_i^t\ ) \times \begin{pmatrix} M_{1i} \\ \vdots \\ M_{ki} \end{pmatrix}$ is like a random sampling process from $\begin{pmatrix} M_{1i} \\ \vdots \\ M_{ki} \end{pmatrix}$.

$Y_i^t = 1$ means $M_{ki}$ is selected and summed into $X_i^{t+1}$.

Therefore, the value of $X_i^{t+1}$ is similar to binomial distribution,

$$P(X_i^{t+1}\ ) = \sum_{n=0}^{N(t)} \binom{N(t)}{k} {E_1}^n {E_{-1}}^{N(t)-n} \quad (12)$$

When N is large enough, $E_{-1} and\ E_1$ is not with great difference, the binomial distribution will approximate to normal distribution,

$$P(X_i^{t+1}\ ) \sim N(u = N(t) \cdot E_1,\ \sigma^2 = N(t) \cdot E_1 \cdot E_{-1}) \quad (13)$$

So the probability of $P(X_i^{t+1} > 0)$ is equal to the difference of the cumulative density function,

$$P(X_i^{t+1} > 0) = \Phi\left(\frac{N(t) - N(t) \cdot E_1}{\sqrt{N(t) \cdot E_1 \cdot E_{-1}}}\right) - \Phi\left(\frac{\frac{N(t)}{2} - N(t) \cdot E_1}{\sqrt{N(t) \cdot E_1 \cdot E_{-1}}}\right)$$

If $N(t)$ is stable, $N(t+1) = N \times P(X_i^{t+1} > 0) = N(t)$.

The value of $N(t)$ is the solution of the function,

$$N(t) = N \cdot \left[\Phi\left(\frac{N(t) - N(t) \cdot E_1}{\sqrt{N(t) \cdot E_1 \cdot E_{-1}}}\right) - \Phi\left(\frac{\frac{N(t)}{2} - N(t) \cdot E_1}{\sqrt{N(t) \cdot E_1 \cdot E_{-1}}}\right)\right] \quad (14)$$

When $E_1$ is within a certain range, $N(t)$ is like a fixed point.

**Proof 3**

It is not difficult to understand that the amount of $Y_i^t$ set is finite, relying on the previous state of the network known as $S(\mathrm{t}-1)$, meaning the state of the network is fixed point or a loop.

If the state of $S(t)$ is a loop, we can begin on any point as $(0) \rightarrow S(1) \rightarrow S(2) \cdots$ $\rightarrow S(t) \rightarrow S(0) \cdots$ , the cycle is $T$. Assuming that cycle $T$ is large enough, and $S(t)$ is unpredictable for an observer, like a series of random systemic states. According to Boltzmann entropy theorem, the entropy for the assembly can be calculated as:

$$S_T = \mathrm{k} \ln T$$

On the other hand, we can also calculate the entropy of the assembly by each neuron by its spike probability $P_i$. According to Shannon entropy theorem,

$$S_P = -\sum_{1}^{i} P_i \log P_i$$

As the amount of $S_T$ is not exactly random, but restricted by the dynamic rules, so $S_T \leq S_P$,

$$\mathrm{k} \ln T \leq -\sum_{1}^{i} P_i \log P_i$$

We can draw some conclusions by analyzing this equation:

(1) In this equation, $T \leq 2^i$. For each neuron, its firing sequence $Y_i^t$ can form a sub-cycle $t_i$. Either at least one $t_i = T$, or $T$ is the least common multiple of all $t_i$.

(2) If the firing sequence $Y_u^t$ of neuron $u$ is limited to pure 0 or 1 (meaning the neuron is under direct stimulation), cycle of the rest neurons will be a new value $T'$. We have already known that $P_i$ will be polarized in this situation, leading to the decrease of Shannon entropy. $T' \leq T$.